# Recognition of Brain Waves of Left and Right Hand Movement Imagery with Portable Electroencephalographs

Zhen Li, Jianjun Xu, Tingshao Zhu

## Abstract

With the development of the modern society, mind control applied to both the recovery of disabled individuals and auxiliary control of normal people has obtained great attention in numerous researches. In our research, we attempt to recognize the brain waves of left and right hand movement imagery with portable electroencephalographs. Considering the inconvenience of wearing traditional multiple-electrode electroencephalographs, we choose Muse to collect data which is a portable headband launched lately with a number of useful functions and channels and it is much easier for the public to use. Additionally, previous researches generally focused on discrimination of EEG of left and right hand movement imagery by using data from C3 and C4 electrodes which locate on the top of the head. However, we choose the gamma wave channels of F7 and F8 and obtain data when subjects imagine their left or right hand to move with their eyeballs rotated in the corresponding direction. With the help of the Common Space Pattern algorithm to extract features of brain waves between left and right hand movement imagery, we make use of the Support Vector Machine to classify different brain waves. Traditionally, the accuracy rate of classification was approximately 90% using the EEG data from C3 and C4 electrode poles; however, the accuracy rate reaches 95.1% by using the gamma wave data from F7 and F8 in our experiment. Finally, we design a plane program in Python where a plane can be controlled to go left or right when users imagine their left or right hand to move. 8 subjects are tested and all of them can control the plane flexibly which reveals that our model can be applied to control hardware which is useful for disabled individuals and normal people.

## Introduction

With the development of the modern society, mind control applied to both the recovery of disabled individuals and auxiliary control of normal people has been hot in numerous researches. In our research, we attempt to recognize the brain waves of left and right hand movement imagery with portable electroencephalographs and we have two main innovations different from the previous related researches. Above all, in traditional researches, the electroencephalographs they used always had many electrodes and they were complex for users to wear. In our research, we will make use of Muse--a portable headband launched by InteraXon Company in 2014 to collect data of brain waves. Muse provides useful functions and channels which are more beneficial for us to build a brain-state-recognition model with CSP and SVM algorithms. Besides, previous researches generally focused on discrimination of EEG of left and right hand movement imagery by using data from C3 and C4—the main controlling parts of movement imagery which locate on the top of the head. On the other hand, we find that eyeballs' movement has influence on brain waves and F7 and F8 electrodes can record the influence well. The places of the corresponding electrodes are shown in Figure 1. Fortunately, with eyeballs rotated in a specified direction, the discrepancy of brain waves of F7 and F8 between left and right hand movement imagery will be more striking. With the help of the Analytic Hierarchy Model, we choose the gamma wave channels of F7 and F8 and obtain data when subjects imagine their left or right hand to move with their eyeballs rotated in the corresponding direction. Then we use the Common Space Pattern algorithm to extract features of brain waves between left and right hand movement imagery and build a Support Vector Machine model to classify different brain waves. Traditionally, the accuracy rate of classification was approximately 90% using the EEG data from C3 and C4 electrode poles; however, the accuracy rate reaches 95.1% by using the gamma wave data from F7 and F8 in our experiment. Furthermore, a display program with a plane is written in Python and users can let the plane left or right in real time by imagining their left or right hand to move with their eyeballs rotated in the corresponding direction. 8 subjects have been tested and all of them think that they can control the plane flexibly. Now that the plane software can be controlled easily, we can apply the same model to control some hardware in the future such as iron-hands, wheelchairs and a number of convenient facilities which are beneficial for both disabled individuals and normal people.

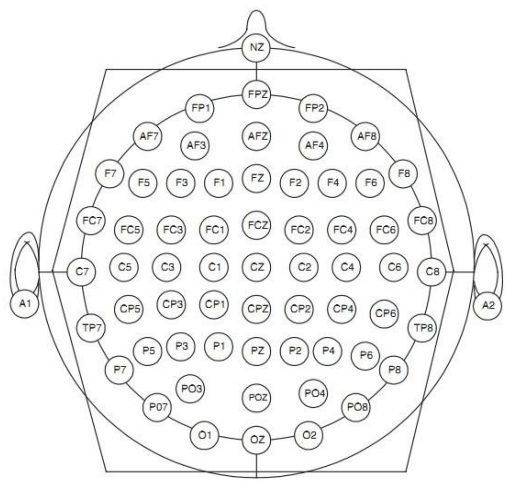

Figure 1

# Related work

As far as we know, a number of researchers have focused on recognition of brain states of left and right hand movement imagery in recent years. Yi Zhang and Mingwei Luo et al.[6] devised an intelligent wheelchair which could be controlled to turn left and right by imagining the left or right hand to move with eyeballs rotated in the specified direction in the mean time. Jingyu Liu and Mingai Li et al.[10] improved the CSP and SVM algorithms in order to increase the recognition accuracy rate of left and right hand movement imagery. Dan Xiao and Kerong He[17] focused on the identification of people through the left hand, right hand, tongue and feet movement imagery. Despite the marvelous contribution they made, their methods had some limitations to some extent. Firstly, the electroencephalographs they used always had many electrode poles which were complex for users to wear. In our research, we make use of Muse--a portable headband with only four electrodes to collect data of brain waves and it should be more likely to be used by public. Furthermore, previous researches generally focused on discrimination of EEG of left and right hand movement imagery by using data from C3 and C4—the main control parts of movement imagery. The traditional recognition accuracy rate was approximately 90% using the EEG data from C3 and C4 electrode poles. However, we made use of data from F7 and F8 and obtained a higher recognition accuracy rate(95.1%) with the help of proper channels and eyeballs' rotation.

# Methods

## 1. The portable electroencephalograph—Muse

In our research, considering the convenience to wear, varieties of data channels and the useful functions, we made use of the portable electroencephalograph—Muse [8] for data collection and tests which was proved to be successful in the following experiment.

Muse was first launched in 2014 by InteraXon Company. Traditionally, electroencephalographs with over 10 electrode poles are usually used in the experiment of dealing with brain signals. These electroencephalographs are generally difficult for subjects to wear and cannot be widely used in people's daily life. Furthermore, traditional electroencephalographs can merely provide limited data source such as EEG, alpha and beta channels. On the other hand, Muse has only 4 electrode poles to measure brain waves of four parts—F7, F8 and the back of ears as reference. The places of the corresponding electrode poles are shown in Figure 2. It is much easier for people to wear instead of using electrode cream before and transmits data collected to the computer via Bluetooth wirelessly. Apart from its convenience to wear, Muse also provides multiple data channels including not only the channels listed above but also gamma wave channels, which will be proved to be more advantageous than others in our experiment. Moreover, Muse provides functions to filter the impure brain waves including the power frequency automatically, which helps increase the signal-noise ratio. Therefore, we use Muse to collect the data of brain waves in the experiment considering its convenience to wear, varieties of data channels and the useful functions.

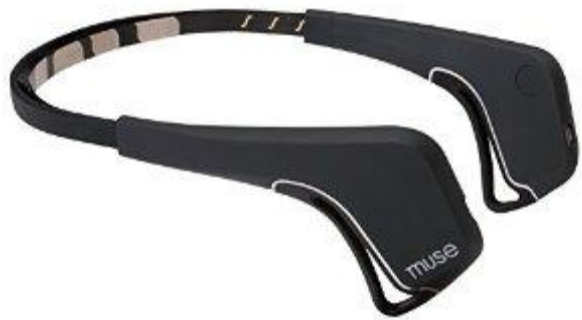

Figure 2 Muse

## 2. Support Vector Machine

In order to choose the proper channels as the data source, we needed to consider the recognition accuracy rate and the number of the channels selected. Firstly, we needed to build a proper classifier to get recognition accuracy rates of different channels for comparison. Like many researchers who study recognition of brain states, we also made use of the Support Vector Machine model[9] to build our classifier.

Due to the complexity and variability of brain waves, the traditional classifiers, which are mainly based on a large sample, lack the stability and accuracy to predict the change of brain waves. As a result, we applied the Support Vector Machine to project the eigenvectors we had extracted to a higher dimensional space, where it was simpler to discriminate the classes. Four kernel functions are mainly used in SVM, and we chose radial basis function for our experiment.

## 3. Feature extraction

After we regarded SVM as the algorithm to build the classifier, the following task for us was to extract the feature of data of different brain states to increase the recognition accuracy rate. According to some related documents, the CSP algorithm[11] can extract the feature of different brain waves effectively. Therefore, we made use of the CSP algorithm to extract the feature of data of different brain states. In the following part, we will first make a short introduction about the CSP algorithm. After we got a series of recognition accuracy rates of different channels using CSP and SVM algorithms, we used the diagram analysis and the Hierarchy Analysis model[12] to choose the most suitable channels.

## 3.1 CSP algorithm

Since brain signals are reflections of numerous active brain sources, the common sources do exist when we imagine our left or right hand to move. Therefore, the discrepancy between the two states is bound to be influenced and the work of extracting the feature is shown to be complicated. In order to separate the related sources from the common sources, we use the Common Space Pattern to calculate spatial filters for detecting ERD/ERS effects. Subsequently, we define the eigenvector and acquire a new data set to train the Support Vector Machine.

## The concrete process of dealing with data with CSP in the experiment

1. We obtained a number of groups of data of left and right hand movement imagery from different subjects.

2. We figured out the spatial filters $W_L$ and $W_R$ for each group. Then the average spatial filters $W_L^A$ and $W_R^A$ could be got.

3. As for a data set $X^T$ with $K$ rows and $N$ columns, we let

$$H_L = W_L^A \times X$$
$$H_R = W_R^A \times X$$

where $H_L \in R^{1 \times K}$ and $H_R \in R^{1 \times K}$.

4. We combined $H_L$ and $H_R$ into a $2K$-dimension eigenvector and it was regarded as the input of SVM.

## 3.2 The choice of channels

As far as we are concerned, the types of data that traditional electroencephalographs provide are always limited. They can only provide data of EEG, alpha waves, beta waves and some other basic data types. On the other hand, Muse can provide us with various types of data including EEG, gamma, beta, alpha, acc, FFT and etc, which contain more information to recognize the brain states. Researchers who have studied the classification of the brain waves between left and right hand movement imagery always use the data from EEG, alpha wave or beta wave channels. In our paper, we built the Analytic Hierarchy Model to choose the optimum channels so as to increase the classification accuracy rate and reduce the complexity of data. F7 and F8 electrode poles were the poles we chose and we intended to choose channels among gamma, beta and alpha wave ones. Thus, the number of the channels we chose might be 2, 4 or 6. As the increase of channels might increase time to run the computer program, we could not choose too many channels to build the classifier. Our purpose was to predict the brain state both accurately and in real time. Therefore, we took the accuracy rate of the prediction, the number of channels and the preknowledge of the types of data into consideration and finally chose the two gamma channels which were attached to the forehead as the data source.

### 3.2.1 Analytic Hierarchy Model

We knew from some psychological papers[6][10] that when a person imagines his left or right hand to move, the beta and gamma waves of the corresponding part of his brain will increase while the ones of the opposite part will decrease. Although the alpha wave has also a good discrimination property between the left and right hand movement imagery, it can only be generated in some specified condition. Therefore, we supposed $c_2$ equaled to 0.5 when alpha channels were used otherwise it equaled to 1. Furthermore, the smaller the number of channels was, the less time it would take to deal with the data and run the computer program so that the classifier was more likely to keep in real time. Thus, we regarded the reciprocal of the number of channels as a factor for us to choose certain channels.

Then, we divided the problem into two hierarchies-the target hierarchy and the factor hierarchy and built the Analytic Hierarchy Model.

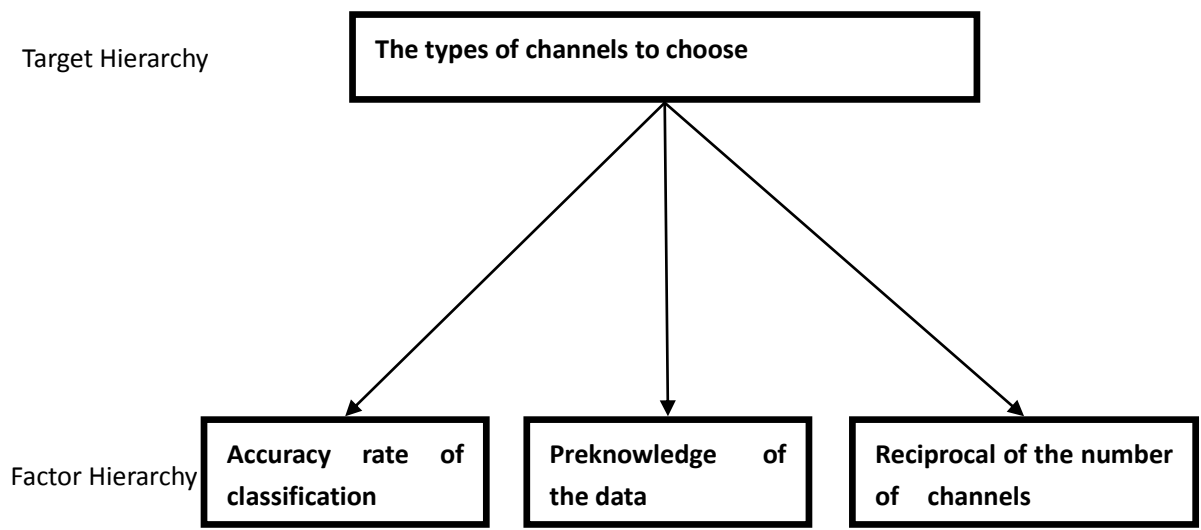

Figure 3 Basic framework of the Analytic Hierachy model

Since the accuracy rate of classification weighed most in the evaluation with no doubt, we gave this factor a rather big weight. Moreover, the number of time to run the computer program differed a little when the number of channels varied; so the factor did not weigh much. Finally, we were not sure that the preknowledge obtained from the Internet suited to our experiment well and we should rely much more on the result of the experiment than the preknowledge; thus, the preknowledge of data only had a slight weight. Therefore, we defined $a_{ij}$ as following:

$$a_{12} = 8$$
$$a_{13} = 3$$
$$a_{23} = 3 \quad (3.1)$$
$$a_{ii} = 1$$
$$a_{ij} = \frac{1}{a_{ji}}, i,j = 1,2,3$$

Then we gave the corresponding comparison matrix

$$A = \begin{bmatrix} 1 & 8 & 3 \\ \frac{1}{8} & 1 & \frac{1}{3} \\ \frac{1}{3} & 3 & 1 \end{bmatrix} \quad (3.2)$$

With standardization, the eigenvector of $\alpha$ was $(0.682, 0.082, 0.236)^T$

Hence, we got the coefficient of the evaluation and let

$$Q = 0.682c1 + 0.082c2 + 0.236c3 \quad (3.3)$$

As the scale of the three factors differed, they should be standardized through being divided by their own maximum number in their range. We could easily know the maximum of the three factors were 1,1,0.5 seperately. From the data we had collected from 8 subjects, we figured out Q for each choice of channels and got the mean values as the following table shows:

|   | **Gamma** | **Beta** | **Alpha** | **Gamma and Beta** |
|---|---|---|---|---|
| **Q** | 0.967 | 0.941 | 0.780 | 0.827 |

Table 1 The value of Q for different channels

As a matter of fact, we could not quantify the importance of the three factors specifically but we had to estimate the relative importance $a_{ij}$ among them initially otherwise the model would not work. Fortunately, the result of the experiment revealed that no matter how the relative importance varied, the channels we should choose were invariable.

Therefore, we chose the two gamma channels of F7 and F8 as the data source which was proved to be successful in the experiment later.

## 4. The plane program

After the proper channels were chosen, we needed to devise suitable software to show people's current brain state of left and right hand movement imagery in real time. A plane program where a plane can be controlled to go left or right when users imagine their left or right hand to move was our final choice.

The purpose of the plane program was to show subjects' brain states of left and right hand movement imagery in real time. Above all, we made use of C++ to write a program to transmit data of brain waves collected by Muse to Python in real time. Subsequently, we transferred the format of the classifiers we made in R Language to the one that was compatible to Python. The output frequency for the gamma wave was 10hz which meant we could obtain 10 rows of data from F7 and F8 per second. As for the SVM model, we regarded 5 rows of data as a whole and got the eigenvector by CSP algorithm as the input of the corresponding classifier. The input was generated every 0.1 second which consisted of data of a subject's current state and 0.5 seconds forward. The plane program was written in Python. If the input was labeled as 1 by the classifier, the plane would go left a bit and if the input was labeled as -1, it would go right a bit. According to the above, the plane would execute a command every 0.1 second to show subjects' brain states of left and right hand movement imagery in real time.

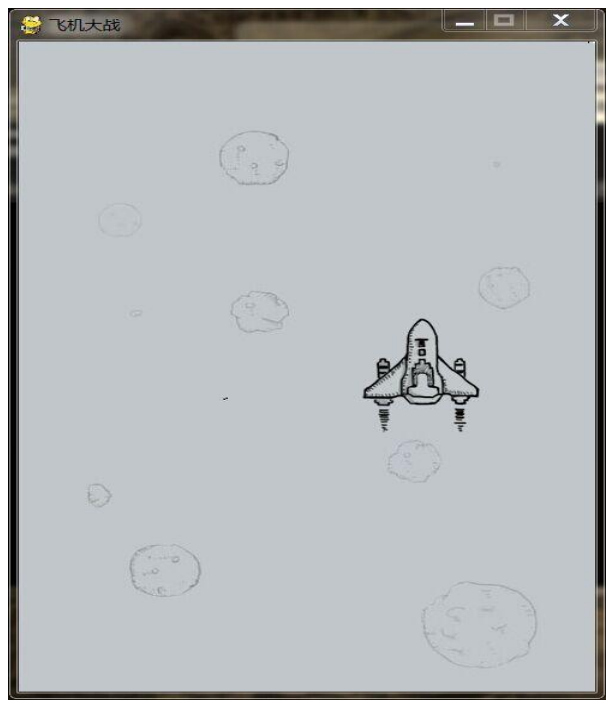

Figure 4 The plane program interface

# 5. The Experiment

After the channels were chosen and the plane software was made, we needed to check out if the feedback system could recognize people's brain state in real time. We designed an experiment to collect the data of brain waves when subjects imagined their left or right hand to move. According to some psychological papers[6], eyeballs' movement has influence on brain waves and F7 and F8 electrodes can record the influence well. If eyeballs are rotated in a specified direction, the discrepancy of brain waves between left and right hand movement imagery will be more striking; therefore, we told the subjects to rotate their eyeballs in the corresponding direction when they imagined their left of right hand to move. Subsequently, a classifier should be made using training samples. The users' data of brain waves would be labeled 1 or -1 by the classifier and the label would be the instruction which controlled the plane to move. When the subject imagined his left hand to move, a plane on the screen would go left and when he imagined his right hand to move, the plane would go right.

## 5.1 Experiment purpose

We need to collect the data of brain waves when a subject imagines his left or right hand to move in order to make a classifier to suit him. With the help of the classifier, the software mentioned above can predict the subject's brain state after the head band is connected to the computer.

## 5.2 Experiment subjects

In the experiment, we collected data of brain waves from 8 adults and tested them with the plane program. The subjects' ages ranged from 19 to 25. Three quarters of the subjects were male and all of the subjects came from China. 25% of the subjects had completed at least some college and the rest of them were undergraduates.

## 5.3 Experiment content

As the power frequency interfered with the brain waves greatly, we had to filter the noise with the help of certain functions in Muse. This was the pretreatment of the experiment. Before the experiment, we divided the subjects into two groups: A and B. At the beginning of the experiment, a subject's forehead should be cleaned with an alcohol cotton ball. Then we helped him wear the headband and made it fit for him. After that, we asked subjects in Group A to imagine their left hand to finish some tasks such as dribbling, waving a badminton racket or taking up a bottle. In the meantime, they should rotate their eyeballs to the left. The process will be done 20 times and 10 seconds for each. Then the subjects may have 2 minutes' rest. After the rest, the subjects were asked to imagine their right hand to finish the same tasks with their eyeballs rotated to the right. The process would be done 20 times and 10 seconds for each likewise. The data of the subjects' brain waves were recorded and saved in csv files. On the other hand, the subjects in Group B were asked to finish the same task by the opposite order. After each experiment, we would make a classifier using the data and figured out the accuracy rate. We would choose 30 of the 40 data sets randomly as the training samples while the rest were the test samples. If the accuracy rate was over 80 percent, we assumed that the result of the test could be accepted. Otherwise, we would analyze the reasons and attempted to improve the experiment procedure.

## 5.4 Time domain Analysis

As we have mentioned before, we chose the two gamma wave channels of F7 and F8 as our data source. The output frequency for the gamma wave was 10hz which meant we could obtain 10 rows of data from the two channels per second. However, despite the filter for the power frequency, not all of the data saved could be effective. According to some related documents[17], we would make use of 5-second data in a 10-second test. We regarded data in different time periods as effective supposedly, made separate classifiers and figured out the mean accuracy rate of each classifier as the following table shows.

|  | 1-5s | 2-6s | 3-7s | 4-8s |
| --- | --- | --- | --- | --- |
| Accuracy rate | 88.3% | 92.5% | 95.1% | 94.2% |

Table 2 The mean accuracy rates of different time periods

As a consequence, we regard the 3-7s data of a 10-second-test as effective.

# Results

## 1. Evaluation values of different channels

We collected data of brain waves from 8 subjects and obtained a series of recognition accuracy rates of different channels. Subsequently, according to the definition of the evaluation function of the Analytic Hierarchy model, we obtained the evaluation values of different channels. The overall evaluation values are shown in Table 3 and the highest(0.967) was of the gamma wave channels which proved that they were the most suitable channels we should choose in the experiment.

| Subjects | Gamma | Beta | Alpha | Gamma and Beta |
|---|---|---|---|---|
| 1 | 0.937 | 0.952 | 0.649 | 0.836 |
| 2 | 0.989 | 0.966 | 0.823 | 0.843 |
| 3 | 0.965 | 0.891 | 0.768 | 0.838 |
| 4 | 0.972 | 0.930 | 0.839 | 0.822 |
| 5 | 0.953 | 0.921 | 0.747 | 0.820 |
| 6 | 0.964 | 0.948 | 0.755 | 0.810 |
| 7 | 0.993 | 0.985 | 0.855 | 0.841 |
| 8 | 0.963 | 0.935 | 0.804 | 0.806 |
| Average | 0.967 | 0.941 | 0.780 | 0.827 |

Table 3 Evaluation values of different channels

## 2. The recognition accuracy rate and mind control effect of the plane program

In the experiment, we tested 8 subjects and the data of them qualified our standard totally. Four order cross validation was used to obtain the recognition accuracy rates so that the results could be regarded as reliable. The overall recognition accuracy rates are shown in Table 4 and the mean accuracy rate was 95.1% which proved that the data we got was relatively pure and effective. After saving the data of a subject, we told him to control the plane program written in Python by imagining his left or right hand to move with his eyeballs rotated in the corresponding direction. After being trained and adjusting, all of them could control the plane to go left or right flexibly which proved that the SVM model could recognize the brain waves of left and right hand movement imagery well.

| Subject | Left | Right | All |
|---|---|---|---|
| 1 | 95.3% | 98.3% | 96.8% |
| 2 | 97.8% | 100% | 98.9% |
| 3 | 94.9% | 89.7% | 92.3% |
| 4 | 95% | 96% | 95.5% |
| 5 | 96.5% | 90.3% | 93.4% |
| 6 | 92.5% | 93.1% | 92.8% |
| 7 | 99.7% | 91.9% | 95.8% |
| 8 | 92.5% | 98.3% | 95.4% |
| Average | 95.5% | 94.7% | 95.1% |

Table 4 The recognition accuracy rates of the SVM classifier

# Discussion

From the results above, users can know about their current brain state of left and right hand movement imagery conveniently with Muse and the recognition accuracy rate is high using data of gamma waves from F7 and F8 with eyeballs rotated in the corresponding direction. Like previous researches, our feedback system can recognize subjects' brain state of left and right hand movement imagery with a relatively high recognition accuracy rate. However, compared to a number of researches, the electroencephalograph with fewer electrodes we use is more portable and more likely to be used widely. Furthermore, we choose the two gamma wave channels of F7 and F8 as the data source instead of EEG channels of C3 and C4 traditionally. With eyeballs rotated in the corresponding direction, the recognition accuracy rate we get is higher than those of previous researches. In our research, we make use of four order cross validation to get the recognition accuracy rates so that the results above can be regarded as believable. Honestly, our feedback system has some limitations. The main limitation is that the high recognition accuracy rate we get has much to do with the instruction of rotating eyeballs in the corresponding direction since the eyes' movements exert much influence on the data of gamma waves from F7 and F8. We are now attempting to devise new algorithms in order to extract obvious features of brain waves from F7 and F8 between left and right hand movement imagery without eyeballs' rotation. Nevertheless, our purpose is to help people control software or hardware flexibly by left and right hand movement imagery. The result of the experiment has shown that users can control the plane program flexibly which means we have achieved our purpose although the eyeballs' rotation is inevitable. As a consequence, the limitation does not have an obvious influence on our aim and the results of the experiment.

## Conclusions

Our research aims at helping people recognize the brain waves of Left and Right Hand Movement Imagery more conveniently and in real time. Considering the inconvenience of wearing traditional multiple-electrodes electroencephalographs, we choose Muse to collect data which is a portable headband launched lately with a number of useful functions and channels and it is much easier for the public to use. Additionally, previous researches generally focused on discrimination of EEG of Left and Right Hand Movement Imagery by using data from C3 and C4 electrodes which locate on the top of the head. However, we choose the gamma wave channels of F7 and F8 and obtain data when subjects imagine their left or right hand to move with their eyeballs rotated in the corresponding direction. With the help of the CSP algorithm to extract features of brain waves between left and right hand movement imagery, we make use of the SVM model to classify different brain waves. Traditionally, the accuracy rate of classification was approximately 90% using the EEG data from C3 and C4 electrode poles; however, the accuracy rate reaches 95.1% by using the gamma wave data from F7 and F8 in our experiment. Finally, 8 subjects are tested and all of them can control the plane flexibly. Therefore in the future, we can apply the same model to control some hardware such as iron-hands, wheelchairs

13and a number of convenient facilities which are beneficial for both disabled individuals and normal people.

# References

[1] Patrick N. Friel, BS. EEG Biofeedback in the Treatment of Attention Deficit Hyperactivity Disorder. *Alternative Medicine Review Volume 12*, Number 2 2007

[2] Xueni Li, Yufeng Wang, Liang Shu, Bomei Gu. A controlled study on the effectiveness of EEG biofeedback on attention deficit hyperactivity disorder. *Chin J Psychiatry,* August 2001, Vol.34, No.3

[3] Yanbing Zhang. The influence of relaxation training on EEG and mood. Achieved from *http://www.docin.com/p-1062964187.html*

[4] Xianfeng Meng. The retrospect and prospect of the relaxation training. *Journal of GzIPF, Vol.24 No.3* 2004

[5] Huiche Science Company. Kether. Achieved from *http://kether.cn/index.html*

[6] Yi Zhang, Mingwei Luo, Yuan Luo, Xiaodong Xu. Intelligent wheelchair human-machine interaction using alpha and beta wave of EEG. *J.Huang Univ. of Sci & Tech. (Natural Science Edition) Vol.41 No.7*  Jul.2013

[7] Weili Ran, Junfei Qiao. BOD Soft-Measuring Approach Based on PCA Time-Delay Neural Network. *Transactions of China Electrotechnical Society Vol.19 No.12*  Dec. 2004

[8] Ariel Garten and Trevor Coleman et al. Muse. Achieved from *http://www.choosemuse.com/*

[9] Xian Hu. Support Vector Machine and R Language. Achieved from *http://blog.jobbole.com/84714/*

[10] Jingyu Liu. Discrimination of EEG between Left and Right Hand Movement Imagery Event. Achieved from *http://www.docin.com/touch/detail.do?id=292949934*

[11] Yinlei Tian, Caihong Zhao. An improved CSP algorithm. DOI:10.3969/j.issn.1672 -0342.2012.02.004

[12] Jiangqi Yuan. Mathematic Model (the third edition). *Higher Education Press*, 2003

[13] Ling Zhang, Ba Zhang. Feedforward Neural Network and integrated algorithms. *Software Press*, 1995;440-448

[14] Max Kuhn. Package 'caret' of R Language. Achieved from *http://caret.r-forge.r-project.org/*

[15] Hong Peng. Electroencephalogram Data Sensing and Awareness. In press

[16] Ming Zhu. Data mining. *USTC Press*

[17] Dan Xiao, Kerong He. The recognition of brain signals of Movement Imagery. *Journal of Jiangxi Blue Sky University, Vol.5 No.2*, June. 2010

[18] Louis V. DiBello, Louis A. Roussos and William Stout. Cognitively Diagnostic Assessment and Psychometric Models. *Handbook of Statistics, Vol. 26* ISSN: 0169-7161  DOI: 10.1016/S0169-7161(06)26031-0